\PassOptionsToPackage{unicode}{hyperref}
\PassOptionsToPackage{hyphens}{url}
\PassOptionsToPackage{dvipsnames,svgnames,x11names}{xcolor}
\documentclass[
]{article}
\usepackage{xcolor}
\usepackage[margin=1in]{geometry}
\usepackage{amsmath,amssymb}
\usepackage{iftex}
\ifPDFTeX
  \usepackage[T1]{fontenc}
  \usepackage[utf8]{inputenc}
  \usepackage{textcomp} 
\else 
  \usepackage{unicode-math} 
  \defaultfontfeatures{Scale=MatchLowercase}
  \defaultfontfeatures[\rmfamily]{Ligatures=TeX,Scale=1}
\fi
\usepackage{lmodern}
\ifPDFTeX\else
\fi
\IfFileExists{upquote.sty}{\usepackage{upquote}}{}
\IfFileExists{microtype.sty}{
  \usepackage[]{microtype}
  \UseMicrotypeSet[protrusion]{basicmath} 
}{}
\makeatletter
\@ifundefined{KOMAClassName}{
  \IfFileExists{parskip.sty}{%
    \usepackage{parskip}
  }{
    \setlength{\parindent}{0pt}
    \setlength{\parskip}{6pt plus 2pt minus 1pt}}
}{
  \KOMAoptions{parskip=half}}
\makeatother
\makeatletter
\ifx\paragraph\undefined\else
  \let\oldparagraph\paragraph
  \renewcommand{\paragraph}{
    \@ifstar
      \xxxParagraphStar
      \xxxParagraphNoStar
  }
  \newcommand{\xxxParagraphStar}[1]{\oldparagraph*{#1}\mbox{}}
  \newcommand{\xxxParagraphNoStar}[1]{\oldparagraph{#1}\mbox{}}
\fi
\ifx\subparagraph\undefined\else
  \let\oldsubparagraph\subparagraph
  \renewcommand{\subparagraph}{
    \@ifstar
      \xxxSubParagraphStar
      \xxxSubParagraphNoStar
  }
  \newcommand{\xxxSubParagraphStar}[1]{\oldsubparagraph*{#1}\mbox{}}
  \newcommand{\xxxSubParagraphNoStar}[1]{\oldsubparagraph{#1}\mbox{}}
\fi
\makeatother

\usepackage{longtable,booktabs,array}
\usepackage{calc} 
\usepackage{etoolbox}
\makeatletter
\patchcmd\longtable{\par}{\if@noskipsec\mbox{}\fi\par}{}{}
\makeatother
\IfFileExists{footnotehyper.sty}{\usepackage{footnotehyper}}{\usepackage{footnote}}
\makesavenoteenv{longtable}
\usepackage{graphicx}
\makeatletter
\newsavebox\pandoc@box
\newcommand*\pandocbounded[1]{
  \sbox\pandoc@box{#1}%
  \Gscale@div\@tempa{\textheight}{\dimexpr\ht\pandoc@box+\dp\pandoc@box\relax}%
  \Gscale@div\@tempb{\linewidth}{\wd\pandoc@box}%
  \ifdim\@tempb\p@<\@tempa\p@\let\@tempa\@tempb\fi
  \ifdim\@tempa\p@<\p@\scalebox{\@tempa}{\usebox\pandoc@box}%
  \else\usebox{\pandoc@box}%
  \fi%
}
\def\fps@figure{htbp}
\makeatother

\NewDocumentCommand\citeproctext{}{}

\makeatletter
 \let\@cite@ofmt\@firstofone
 \def\@biblabel#1{}
 \def\@cite#1#2{{#1\if@tempswa , #2\fi}}
\makeatother
\newlength{\cslhangindent}
\newlength{\csllabelwidth}
\newenvironment{CSLReferences}[2] 
 {\begin{list}{}{%
  \setlength{\itemindent}{0pt}
  \setlength{\leftmargin}{0pt}
  \setlength{\parsep}{0pt}
  \ifodd #1
   \setlength{\leftmargin}{\cslhangindent}
   \setlength{\itemindent}{-1\cslhangindent}
  \fi
  \setlength{\itemsep}{#2\baselineskip}}}
 {\end{list}}
\usepackage{calc}

\usepackage{tabularx}
\usepackage{float}
\usepackage{longtable}
\usepackage{array}
\usepackage{hyperref}
\makeatletter
\@ifpackageloaded{caption}{}{\usepackage{caption}}
\AtBeginDocument{%
\ifdefined\contentsname
  \renewcommand*\contentsname{Table of contents}
\else
  \newcommand\contentsname{Table of contents}
\fi
\ifdefined\listfigurename
  \renewcommand*\listfigurename{List of Figures}
\else
  \newcommand\listfigurename{List of Figures}
\fi
\ifdefined\listtablename
  \renewcommand*\listtablename{List of Tables}
\else
  \newcommand\listtablename{List of Tables}
\fi
\ifdefined\figurename
  \renewcommand*\figurename{Figure}
\else
  \newcommand\figurename{Figure}
\fi
\ifdefined\tablename
  \renewcommand*\tablename{Table}
\else
  \newcommand\tablename{Table}
\fi
}
\@ifpackageloaded{float}{}{\usepackage{float}}
\floatstyle{ruled}
\@ifundefined{c@chapter}{\newfloat{codelisting}{h}{lop}}{\newfloat{codelisting}{h}{lop}[chapter]}
\floatname{codelisting}{Listing}

\makeatother
\makeatletter
\@ifpackageloaded{caption}{}{\usepackage{caption}}
\@ifpackageloaded{subcaption}{}{\usepackage{subcaption}}
\makeatother
\usepackage{bookmark}
\IfFileExists{xurl.sty}{\usepackage{xurl}}{} 
\hypersetup{
  pdftitle={Sri Lanka Dengue Dashboard: A Web-Based Decision Support Tool for Dengue Surveillance in Sri Lanka},
  pdfauthor={Y.M. Amali P. Rajapaksha Department of Statistics, Faculty of Applied Sciences Unversity of Sri Jayewardenepura, Sri Lanka},
  colorlinks=true,
  linkcolor={blue},
  filecolor={Maroon},
  citecolor={Blue},
  urlcolor={Blue},
  pdfcreator={LaTeX via pandoc}}

\title{Sri Lanka Dengue Dashboard: A Web-Based Decision Support Tool for
Dengue Surveillance in Sri Lanka}
\author{Y.M. Amali P. Rajapaksha \\ Department of Statistics, Faculty of
Applied Sciences \\ Unversity of Sri Jayewardenepura, Sri Lanka}
\date{}
\begin{document}
\maketitle

\subsection{\texorpdfstring{\textbf{Abstract}}{Abstract}}\label{abstract}

These days, Sri Lanka is facing a severe dengue outbreak. In this
situation, it is very useful to provide an overall understanding of Sri
Lanka's historical dengue trends as well as the current dengue
situation. A dashboard is an easy and effective way to convey a large
amount of information to the public. Therefore, the \textbf{Sri Lanka
Dengue Dashboard} is an initial attempt to provide comprehensive
information about the dengue situation in Sri Lanka, together with the
relationship between climate factors and dengue cases.\\
\strut \\
\textbf{Keywords} : Data visualization, Heat map, Climate-dengue
relationship, Tree map, Time series analysis, Quato, Dashboard, Dengue
fever

\section{1 Introduction}\label{introduction}

Dengue is endemic in more than 100 tropical and subtropical countries
worldwide (Paz-Bailey et al., 2024). It is transmitted to humans through
the bites of infected Aedes mosquitoes. Currently, there are no specific
medicines for dengue fever, and a person can be infected with dengue
more than once during his or her lifetime (World Health Organization,
2025). Hence, prevention is better than cure in this case.

Asian countries such as Bangladesh, India, Thailand, and Sri Lanka also
face dengue outbreaks. Currently, Sri Lanka is experiencing an
increasing trend in dengue cases. Therefore, having a better
understanding of past trends, seasonal patterns, and the current dengue
situation may help to prevent and control the disease.

In this context, an interactive dashboard is an effective way to
communicate a large amount of information to the public, health
professionals, and decision-makers in a simple and accessible manner.
There are many dashboards designed to visualize the global and local
dengue situation. Among those, I studied six different dashboards
covering the worldwide, Asian, South-East Asian regional, and
country-level dengue situations in selected South-East Asian countries.
Based on the insights gained from these dashboards, I developed a
dashboard specifically for Sri Lanka. This dashboard provides an
interactive overview of the dengue situation in Sri Lanka using the
latest available surveillance data. Built with R using the Quarto
Dashboard framework, it presents recent dengue trends, district-level
analyses, historical patterns, comparisons with the global dengue
situation, and explores the relationship between dengue transmission and
key climate variables. The dashboard offers valuable insights for
monitoring, understanding, and supporting evidence-based decision-making
on dengue dynamics in Sri Lanka.

The remaining sections of this paper are organized as follows. Section 2
presents a brief literature review based on six dashboards developed for
different contexts using dengue surveillance data. Section 3 describes
the methodologies used in developing the dashboard. Section 4 presents
the results and the possible insights obtained from the dashboard.
Finally, Section 5 provides the discussion and concluding remarks.

\section{2 Literature Review}\label{literature-review}

In this literature review, six dengue-related dashboards were
considered. These dashboards were developed for different geographical
areas. The review focuses on the information presented, visualization
techniques, and key features of each dashboard. A summary of the studied
dashboards is presented in Table~\ref{tbl-dashboard-summary}.

\begin{longtable}{|c|p{3.3cm}|p{2.8cm}|p{4.7cm}|p{2.8cm}|}

\caption{\label{tbl-dashboard-summary}Summary of dengue surveillance dashboards considered in the literature review}

\tabularnewline

\hline
\textbf{No.} & \textbf{Dashboard Name} & \textbf{Geographical Area Covered} & \textbf{Data Included in the Dashboard} & \textbf{Citation} \\
\hline
\endfirsthead

\multicolumn{5}{c}%
{{\bfseries Table \thetable\ Continued from previous page}} \\
\hline
\textbf{No.} & \textbf{Dashboard Name} & \textbf{Geographical Area Covered} & \textbf{Data Included in the Dashboard} & \textbf{Citation} \\
\hline
\endhead

\hline
\multicolumn{5}{r}{{Continued on next page}}\\
\endfoot

\hline
\endlastfoot

1 & Global Dengue Surveillance Dashboard & Worldwide &  Total cases, Confirmed cases, Severe cases, Deaths, Total cases per 100 000, Geographic distribution, Frequencies by serotype, Global dengue transmission range estimates (2023)  & (WorldHealth Organization, 2026) 
 \\
\hline

2 & Asia Dengue Voice and Action: Dengue Dashboard  & Asia Region & Weekly Dengue Cases, Total Dengue Cases, Total Dengue Deaths
& (AsianDengue Voice and Action, 2026)\\
\hline

3 & Dengue Dashboard: South-East Asia Region & South-East Asia Region & Total Dengue Cases, Confirmed Dengue Cases, Dengue Deaths, Gender \& Age overview
, Severity overview
, Serotyping, Country profile, Clinical Management & (WorldHealth Organization Regional Office for South-East Asia, 2026) \\
\hline

4 & Dengue Dynamic Dashboard for Bangladesh & Bangladesh & Weekly Cases, Weekly Deaths, Cumulative Cases, Cumulative Deaths, Dengue cases for last 24 hours, Dengue deaths for last 24 hours, Division \& City corporation cases of last 24 hours, Division \& City corporation deaths of last 24 hours, Age group distribution of affected cases of last 24 hours, Gender distribution of affected cases of last 24 hours & (DirectorateGeneral of Health Services (DGHS), Bangladesh, 2026)\\
\hline

5 & WHO Health Emergencies Dashboard: Outbreaks - Afghanistan & Afghanistan & Suspected cases, Geographical distribution of Suspected Dengue Fever Cases by District, Dengue fever cases by gender, Dengue fever cases by age group, Dengue fever deaths by gender and age & (WorldHealth Organization, 2026)
 \\
\hline

6 & Dengue Daily Data Dashboard: Karnataka & Karnataka & Daily Counts for All Districts, Weekly Trend, Bi-Weekly Trend, Monthly Trend & (InternationalCentre for Theoretical Sciences (ICTS), 2026)
 \\
\hline

\end{longtable}

Table~\ref{tbl-dashboard-features} numbers correspond to those listed in
Table~\ref{tbl-dashboard-summary}. The comparison focuses on the number
of tabs, background theme, suitability for a single-screen layout, data
availability, and the visualization techniques used in each dashboard.

\begin{longtable}{|c|c|p{2cm}|p{2cm}|p{2.5cm}|p{4.7cm}|}

\caption{\label{tbl-dashboard-features}Comparison of visualization techniques and key features of the reviewed dengue dashboards}

\tabularnewline

\hline
\textbf{No.} &
\textbf{No. of Tabs} &
\textbf{Background Theme} &
\textbf{Single-Screen Layout} &
\textbf{Public Data Availability} &
\textbf{Visualization Techniques} \\
\hline
\endfirsthead

\multicolumn{6}{c}{{\bfseries Table \thetable\ Continued from previous page}}\\
\hline
\textbf{No.} &
\textbf{No. of Tabs} &
\textbf{Background Theme} &
\textbf{Single-Screen Layout} &
\textbf{Public Data Availability} &
\textbf{Visualization Techniques} \\
\hline
\endhead

\hline
\multicolumn{6}{r}{{Continued on next page}}\\
\endfoot

\hline
\endlastfoot

1 & 7 & light & No & Available & KPIs, Choropleth Map, Bar Chart, Area chart, Stacked Bar Chart,   \\
\hline

2 & 1 & light & No & Not Available & Choropleth Map, Line Chart \\
\hline

3 & 6 & light & No & Available & KPIs, Choropleth Map, Bar Chart, Area chart, Stacked Bar Chart, Tree Map, Heatmap, Line Chart \\
\hline

4 & 1 & light & No & Not Available & KPIs, Bar Chart, Line Chart, Pie Chart,  \\
\hline

5 & 1 & light & No & Not Available & Choropleth Map, Bar Chart, Donut Chart \\
\hline

6 & 6 & light & No & Available & Choropleth Map, Line Chart \\
\hline

\end{longtable}

Among these dashboards, tree maps and heatmaps were used in only one
dashboard, while line charts and bar charts were used most frequently.

\section{3 Data}\label{data}

All local data up to December 2025 used in this dashboard were obtained
from the weekly epidemiological reports published by the Epidemiology
Unit, Ministry of Health, Sri Lanka, through the \texttt{denguedatahub}
R package (T. Talagala, 2026). Data for 2026 were obtained from the
weekly epidemiological reports (Epidemiology Unit, 2026) published by
the Ministry of Health, Sri Lanka, and the Weekly Dengue Updates
(National Dengue Control Unit, 2026) published by the Ministry of
Health, Sri Lanka . These data were extracted through web scraping using
the \texttt{convert\_slwer\_to\_tidy()} function available in the
\texttt{denguedatahub} package (T. Talagala, 2026). The global data used
in this dashboard were obtained from the World Health Organization's
\href{https://worldhealthorg.shinyapps.io/dengue_global/}{Global Dengue
Surveillance Dashboard} (World Health Organization, 2026a). The climate
data used in this study were downloaded from the
\href{https://power.larc.nasa.gov/data-access-viewer/}{NASA POWER Data
Access Viewer} (NASA Langley Research Center, 2026).

\section{4 Methodology}\label{methodology}

Here, several visualization techniques have been used for different
purposes. To explore new visualization techniques that can be applied
for disease surveillance and epidemiological analysis, the Sri Lanka
COVID-19 Dashboard (T. S. Talagala \& Shashikala, 2022) was referred to.
To get an overall idea of the methodologies, see
Table~\ref{tbl-methods}.

\begin{longtable}{|p{9cm}|p{6cm}|}

\caption{\label{tbl-methods}Summary of methodologies used in the dashboard}

\tabularnewline

\hline
\textbf{Information Presented} & \textbf{Visualization Technique(s)} \\
\hline
\endfirsthead

\multicolumn{2}{c}%
{{\bfseries Table \thetable\ Continued from previous page}}\\
\hline
\textbf{Information Presented} & \textbf{Visualization Technique(s)} \\
\hline
\endhead

\hline
\multicolumn{2}{r}{{Continued on next page}}\\
\endfoot

\hline
\endlastfoot
Key Information about the Present Dengue Situation in Sri Lanka & KPIs \\
\hline
Distribution of Total Dengue Cases in Sri Lanka in 2026 & Choropleth Map \\
\hline
Cumulative Dengue Deaths in Sri Lanka in 2026 & Choropleth Map \\
\hline
Anomalous Dengue Behaviour in the Current Period Compared with the Previous Two Years in Sri Lanka & Time Series Plot \\
\hline
Last Week's Dengue Situation Analysis & Choropleth Maps \\
\hline
District-wise Weekly Dengue Case Analysis Over Time until 2025 & Faceted and Individual Time Series Plots, Faceted and Individual Seasonal Plots \\
\hline
Distribution of Dengue Patients by District and Year until 2025 & Tree Map, Faceted Choropleth Map, Heatmap \\
\hline
Relationship Between Climatic Variables and Dengue Cases & Multi-panel Time Series Plot, Scatterplots \\
\hline
Global Distribution of Dengue Patients/Cases in 2025 & Choropleth Map \\
\hline
Sri Lanka Compared with the Top 10 Dengue-Affected Countries Worldwide and Other South-East Asian countries in 2025 & Stacked Bar Chart \\
\hline

\end{longtable}

One of the additional analyses included in this dashboard is the
climate-dengue relationship analysis. In addition to time series plots
(Figure~\ref{fig-tab6}(a)) and scatter plots (Figure~\ref{fig-tab6}(b))
of climatic variables vs.~dengue cases, I explored the potential lag
effect using two-week and four-week lag scatter plots
(Figure~\ref{fig-tab6}(c) , Figure~\ref{fig-tab6}(d)). Dengue
transmission does not respond immediately to climatic changes such as
rainfall and humidity. Instead, favourable climatic conditions may
enhance mosquito breeding and contribute to increased dengue cases after
a certain time lag. Based on this idea, I explored the lag relationship
between climatic factors and dengue cases. Furthermore, to obtain a
comprehensive understanding of the relationship between climatic
variables and dengue cases, the Detrended Cross-Correlation Coefficient
(\(\rho_{DCCA}\)) was calculated across different time scales (Figueredo
et al., 2023). This method is used to study the relationship between
multiple time series that change over time, even when the data are not
stationary (Zebende et al., 2018). The calculated \(\rho_{DCCA}\) values
were plotted over different time scales to investigate how climatic
variables influence dengue occurrence across short, medium and long-term
periods.

In this dashboard, several colour palettes were used ensuring that the
visualizations are colour-blind friendly. These include the Viridis
palette from the viridis package (Garnier et al., 2018), the Paired
palette from the ColorBrewer collection (Harrower \& Brewer, 2003), and
the Magma palette from the Viridis colour map family (Garnier et al.,
2018). In addition, some colour themes were manually defined according
to the context of the visualizations.

In this dashboard, multiple visualization techniques have sometimes been
used for the same data. Although the data are the same, each
visualization provides different insights that can be directly obtained
from the data. Therefore, using multiple plots for the same data helps
explore different aspects of the actual situation more comprehensively.

\section{5 Results}\label{results}

The dashboard is organized into 8 navigation tabs, including Overview,
Last Week Dengue Situation Analysis, District-wise Time Series Trends,
District-wise Seasonal Plots, Dengue Patients Distribution,
Climate-Dengue Relationship Analysis, Global Comparison, and About. Each
tab contains interactive visualizations and analytical outputs.
Table~\ref{tbl-tabs} provides a description of each tab and summarizes
its key content.

\begin{longtable}{|p{2cm}|p{13cm}|}

\caption{\label{tbl-tabs}Description of Dashboard Tabs}

\tabularnewline

\hline
\textbf{Tab No.} & \textbf{Description} \\
\hline
\endfirsthead

\multicolumn{2}{c}%
{{\bfseries Table \thetable\ Continued from previous page}}\\
\hline
\textbf{Tab No.} & \textbf{Description} \\
\hline
\endhead

\hline
\multicolumn{2}{r}{{Continued on next page}}\\
\endfoot

\hline
\endlastfoot
Tab 1 & This provides the latest information about the dengue epidemic in Sri Lanka. It gives a brief overview of the current dengue situation in Sri Lanka. \\
\hline
Tab 2 & This compares the dengue situation in the 29th week of this year with the same week of the previous year. This is the latest data we have. \\
\hline
Tab 3 & There are two sub-tabs. The first contains faceted time series plots of weekly dengue cases from 2006 to 2025 for each district separately. The second tab allows users to explore small changes in dengue cases for each district individually using dygraphs. \\
\hline
Tab 4 & There are two sub-tabs. The first contains faceted seasonal plots of weekly dengue cases from 2006 to 2025 for each district separately. The second tab allows users to explore small changes in seasonal patterns for each district individually using Plotly graphs. \\
\hline
Tab 5 & There are three sub-tabs. The first contains a treemap showing the distribution of total dengue patients in 2025. The second and third sub-tabs contain the distribution of dengue patients over the last 20 years by district and year.   \\
\hline
Tab 6 & This section contains five sub-tabs and provides a comprehensive analysis of the relationship between climatic variables and dengue cases. The first four sub-tabs present time series plots and scatter plots to explore temporal patterns and associations between climatic variables and dengue incidence. The fifth sub-tab presents a map displaying the calculated raw DCCA values across different time scales.\\
\hline
Tab 7 & This section has three sub-tabs. The first shows the distribution of dengue patients worldwide in 2025. The second and third sub-tabs compare the dengue situation in Sri Lanka with the global and South-East Asia region levels, respectively. \\
\hline
Tab 8 & This contains detailed information about the dashboard, data sources, references, and other related information. \\
\hline

\end{longtable}

\begin{figure}[H]

\centering{

\includegraphics[width=5.09in,height=\textheight,keepaspectratio]{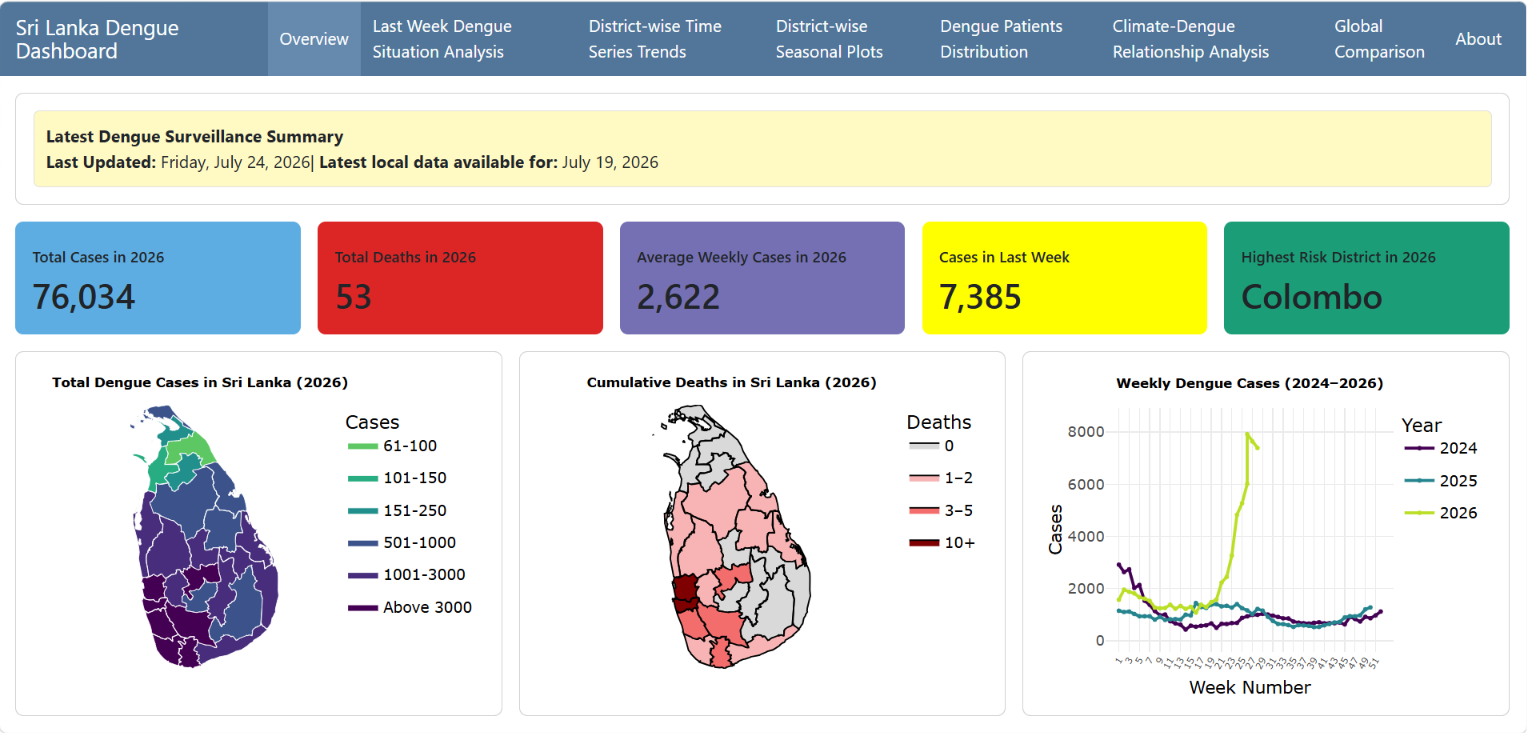}

}

\caption{\label{fig-tab1}Screenshot of Tab 1 (Overview)}

\end{figure}%

According to Figure~\ref{fig-tab1}, a rapid increase in dengue cases can
be observed in 2026 from around the 21st week compared with 2024 and
2025. Public sources suggest that this increase may be associated with
enhanced Aedes mosquito breeding due to favourable climatic conditions,
post-flood environmental changes following Cyclone Ditwah,
urbanization-related breeding habitats, and ongoing dengue virus
transmission dynamics.

\begin{figure}[H]

\centering{

\includegraphics[width=5.11in,height=\textheight,keepaspectratio]{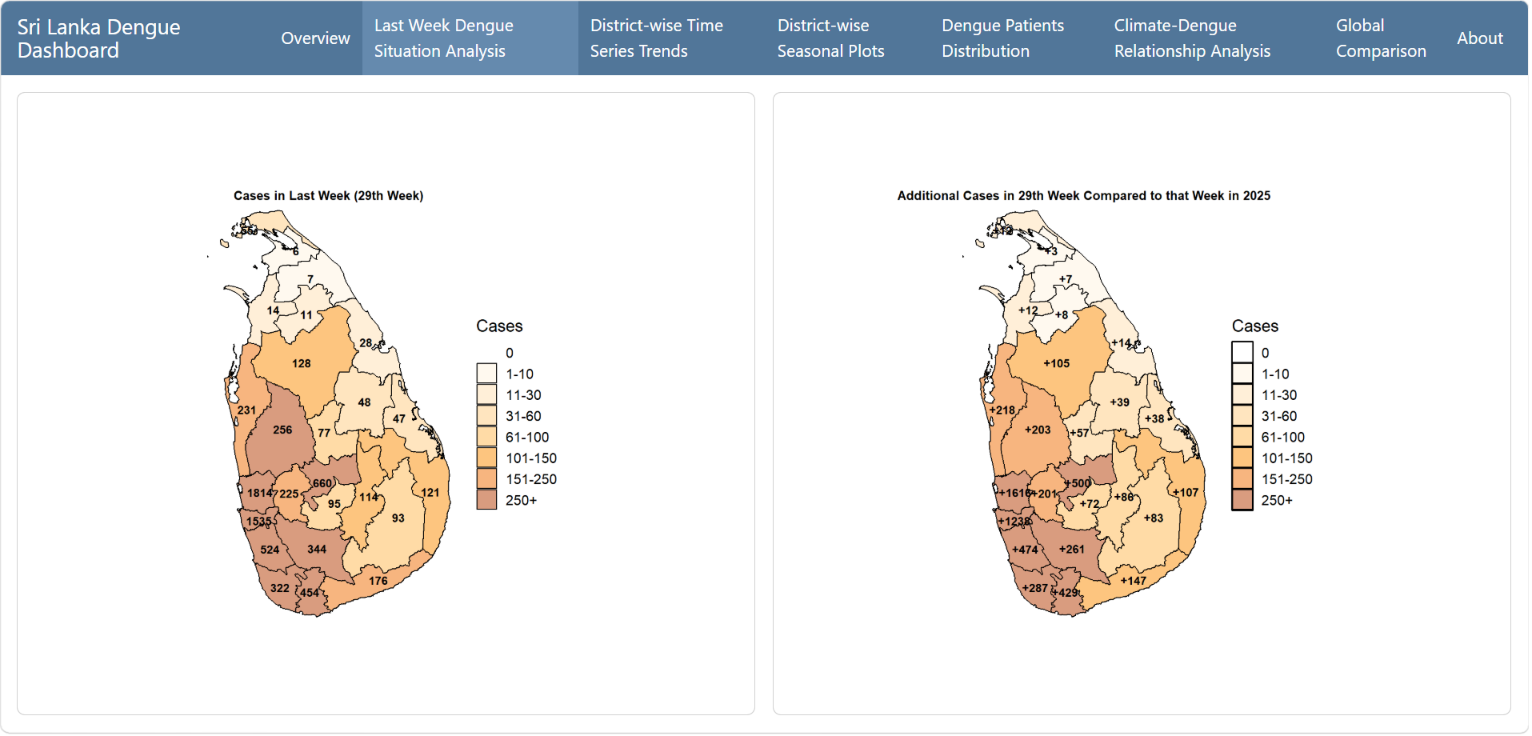}

}

\caption{\label{fig-tab2}Screenshot of Tab 2 (Last Week Dengue Situation
Analysis)}

\end{figure}%

Figure~\ref{fig-tab2} also proves the severity of the situation Sri
Lanka faces today. Approximately, the number of cases reported in the
29th week this year is nearly twice that of the same week last year.\\

\begin{figure}[H]

\begin{minipage}[t]{0.50\linewidth}

\centering{

\includegraphics[width=0.98\linewidth,height=\textheight,keepaspectratio]{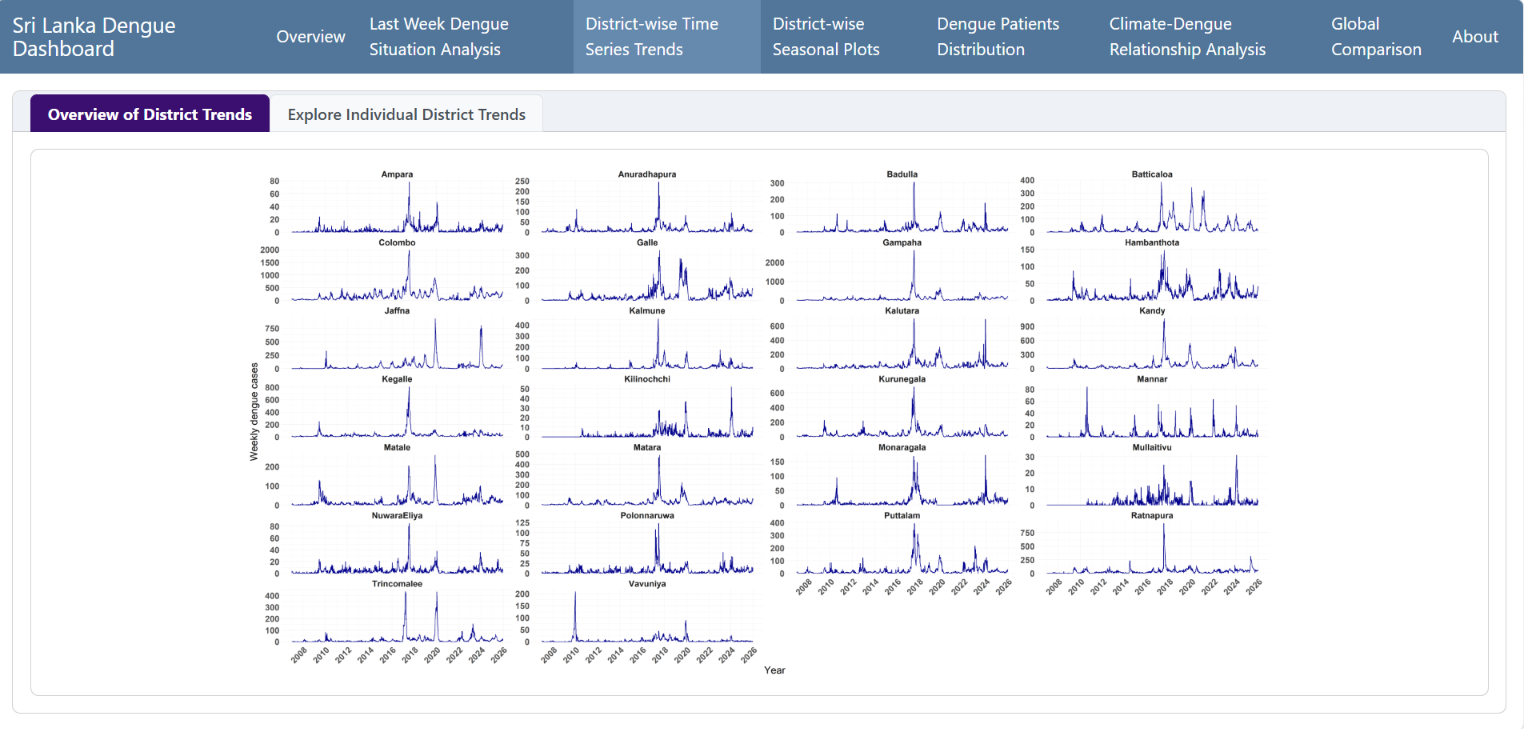}

}

\subcaption{\label{fig-tab3-1}}

\end{minipage}%
\begin{minipage}[t]{0.50\linewidth}

\centering{

\includegraphics[width=0.98\linewidth,height=\textheight,keepaspectratio]{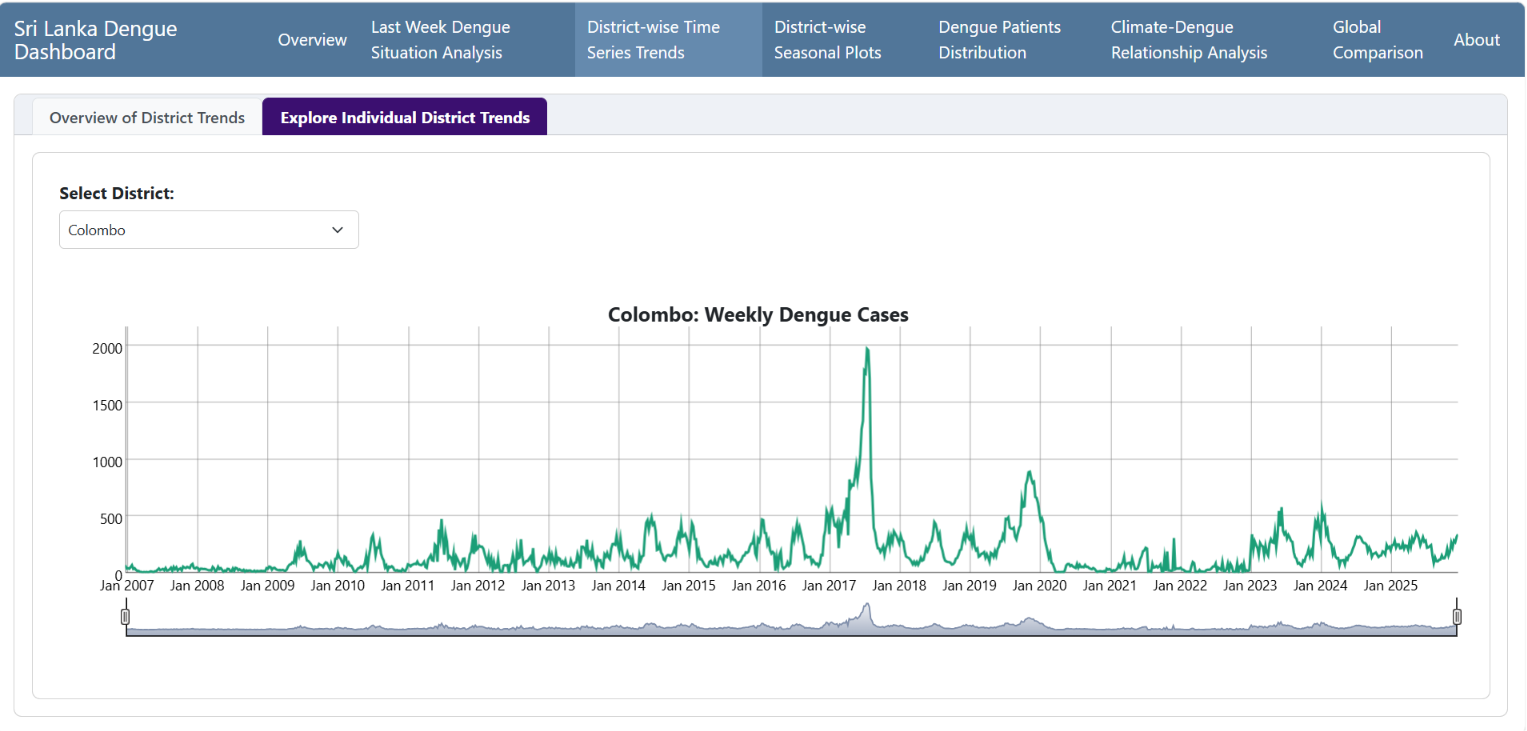}

}

\subcaption{\label{fig-tab3-2}}

\end{minipage}%

\caption{\label{fig-tab3}Screenshot of Tab 3 (District-wise Time Series
Trends)}

\end{figure}%

Figure~\ref{fig-tab3}(a) shows that many districts experienced notable
peaks in dengue cases during 2017, 2019, and 2023. Among these, the 2017
outbreak was the most severe, with consistently high case counts across
nearly all districts. According to published reports, Sri Lanka recorded
its largest dengue outbreak in 2017, with 186,101 suspected cases and
440 deaths nationwide. Preliminary laboratory investigations identified
Dengue virus serotype 2 (DENV-2) as the predominant circulating strain
associated with this outbreak. Heavy monsoon rains and flooding also
contributed to the outbreak by creating favourable breeding conditions
for Aedes mosquitoes through standing water and inadequate waste
management, leading to increased transmission, particularly in urban and
suburban areas.

Dengue transmission in Sri Lanka shows a clear seasonal pattern, with
two annual peaks corresponding to the Southwest (May to September) and
Northeast (October to January) monsoon periods. However, this pattern is
not observed uniformly across all districts.

\begin{figure}[H]

\begin{minipage}[t]{0.50\linewidth}

\centering{

\includegraphics[width=0.98\linewidth,height=\textheight,keepaspectratio]{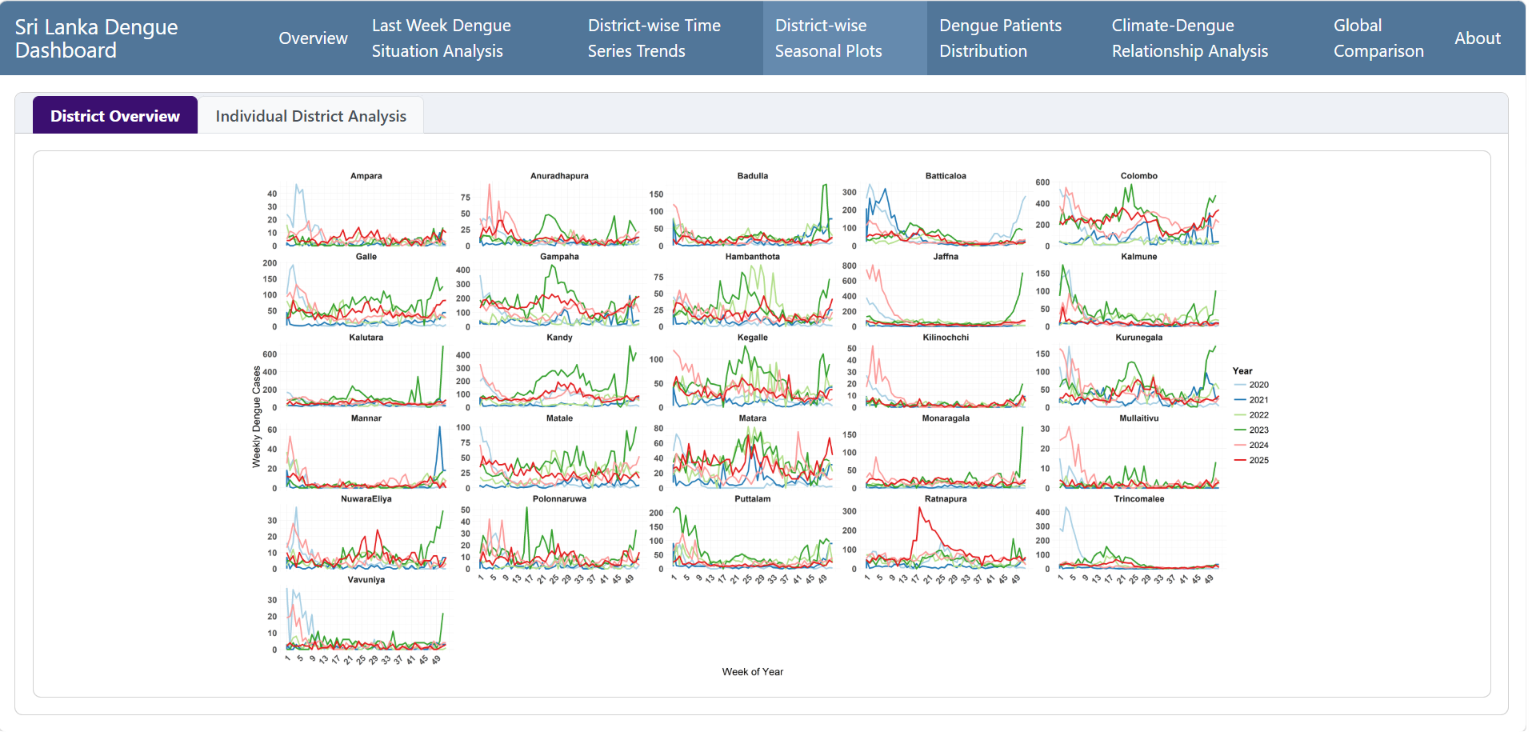}

}

\subcaption{\label{fig-tab4-1}}

\end{minipage}%
\begin{minipage}[t]{0.50\linewidth}

\centering{

\includegraphics[width=0.98\linewidth,height=\textheight,keepaspectratio]{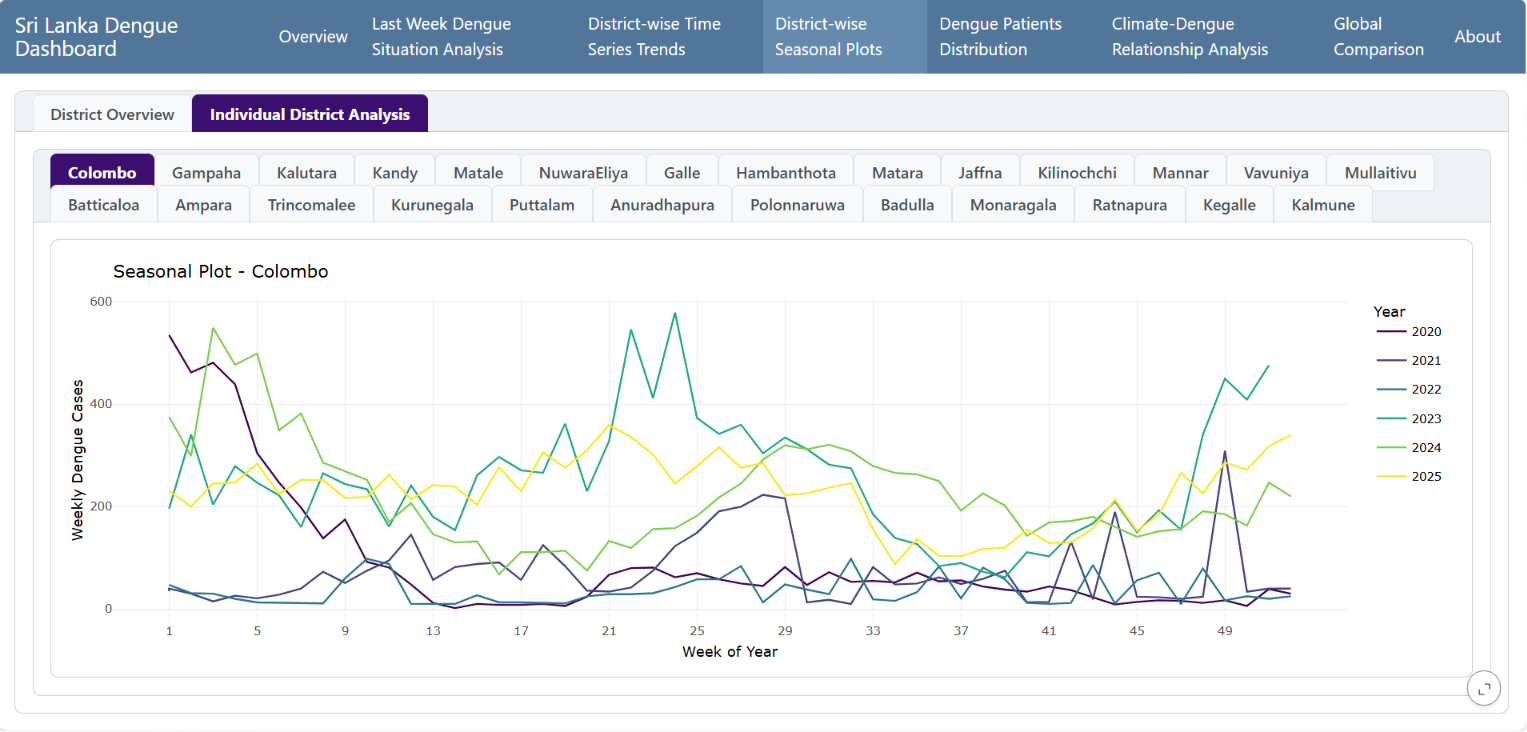}

}

\subcaption{\label{fig-tab4-2}}

\end{minipage}%

\caption{\label{fig-tab4}Screenshot of Tab 4 (District-wise Seasonal
Plots)}

\end{figure}%

The Southwest monsoon mainly affects districts such as Colombo, Gampaha,
Kalutara, Galle, Matara, Hambantota, Kandy, Matale, Nuwara Eliya,
Ratnapura, Kegalle, Puttalam, and Kurunegala. In contrast, districts
such as Trincomalee, Batticaloa, Ampara, Jaffna, Kilinochchi, Mannar,
Mullaitivu, Vavuniya, Badulla, and Monaragala are not significantly
affected by the Southwest monsoon. This difference is clearly visible in
Figure~\ref{fig-tab4}(a). While the Southwest monsoon affected districts
show annual peaks in dengue cases during the period from May to
September, districts such as Trincomalee, Batticaloa, Ampara, Jaffna,
Kilinochchi, Mannar, Vavuniya, Badulla, and Monaragala do not exhibit a
similar seasonal peak during this period.

\begin{figure}[H]

\begin{minipage}[t]{0.50\linewidth}

\centering{

\includegraphics[width=0.98\linewidth,height=\textheight,keepaspectratio]{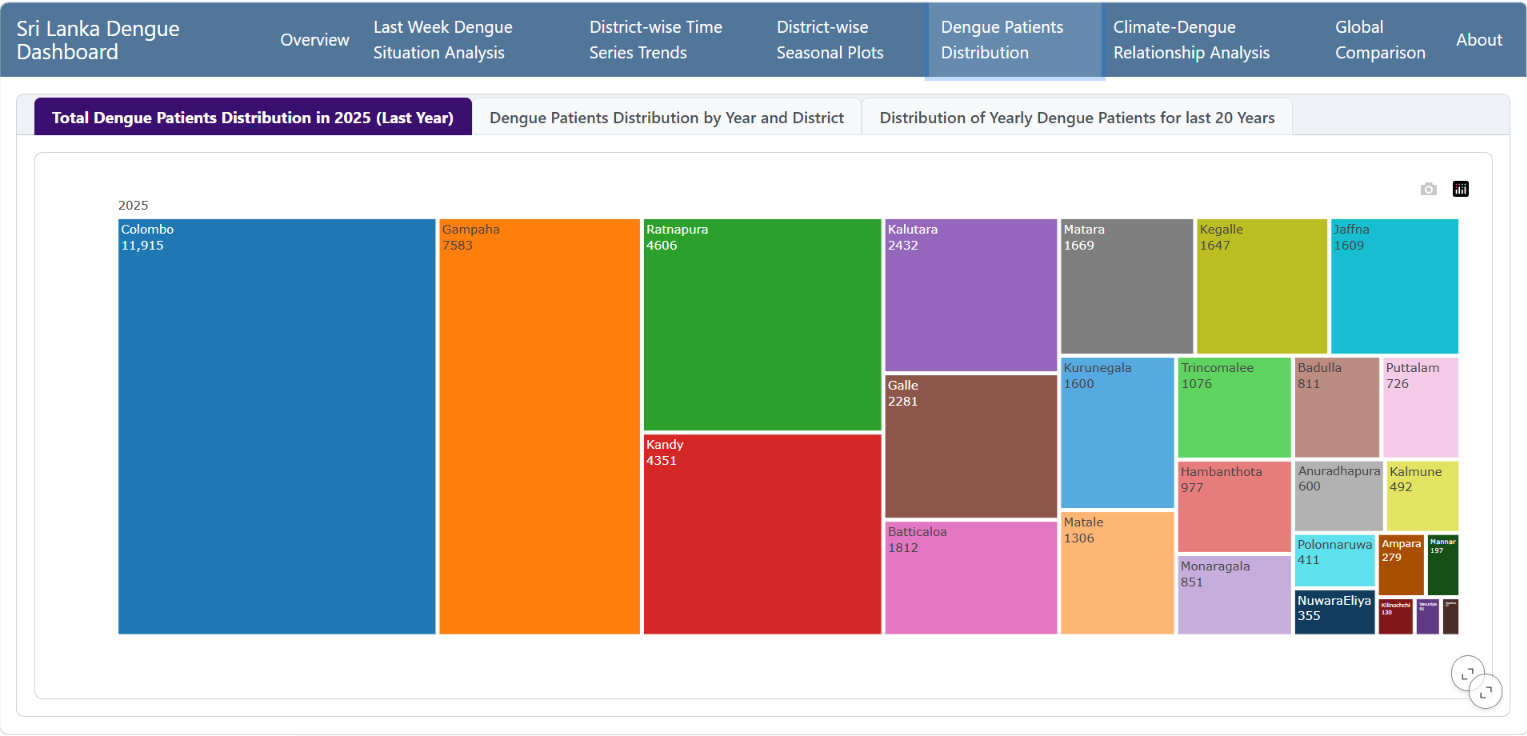}

}

\subcaption{\label{fig-tab5-1}}

\end{minipage}%
\begin{minipage}[t]{0.50\linewidth}

\centering{

\includegraphics[width=0.98\linewidth,height=\textheight,keepaspectratio]{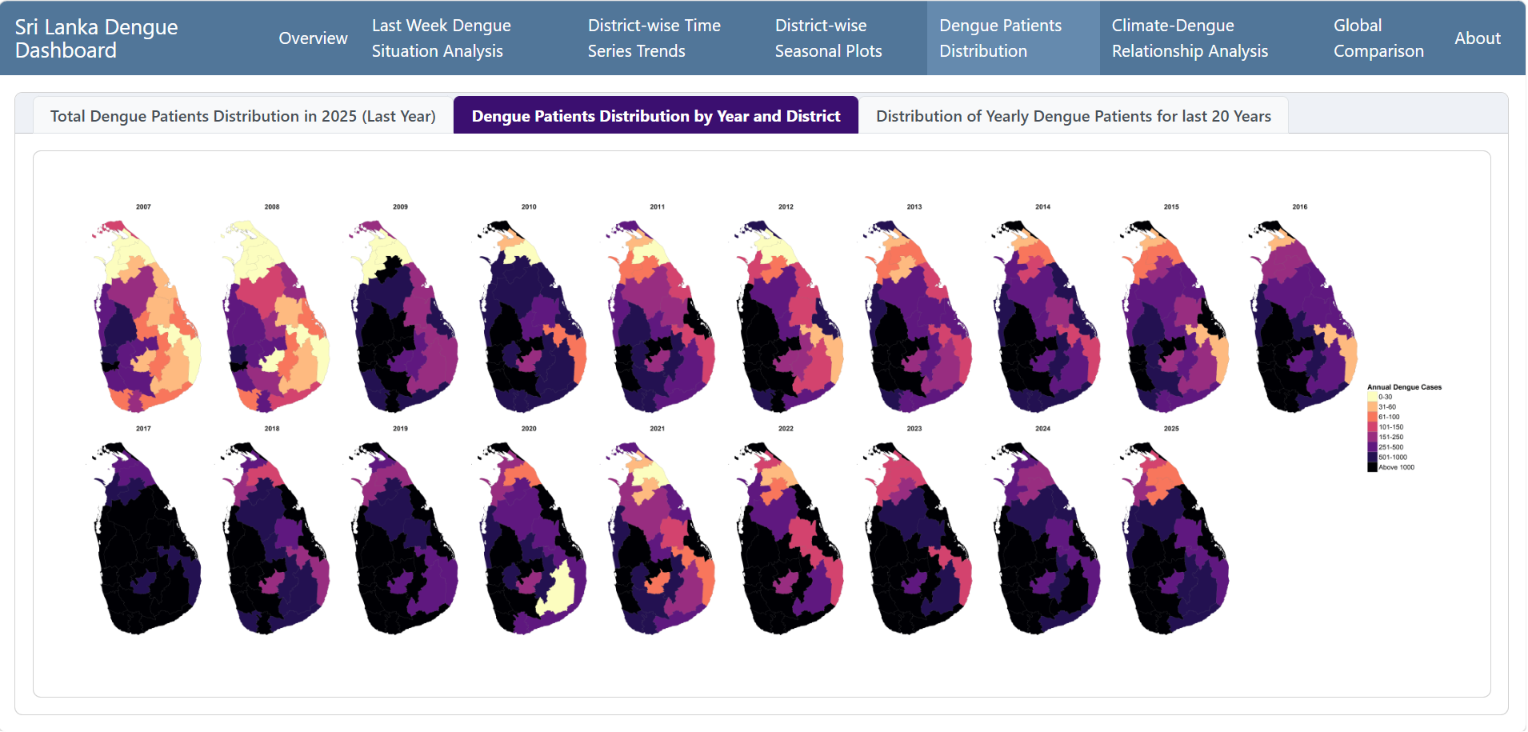}

}

\subcaption{\label{fig-tab5-2}}

\end{minipage}%
\newline
\begin{minipage}[t]{0.50\linewidth}

\centering{

\includegraphics[width=0.98\linewidth,height=\textheight,keepaspectratio]{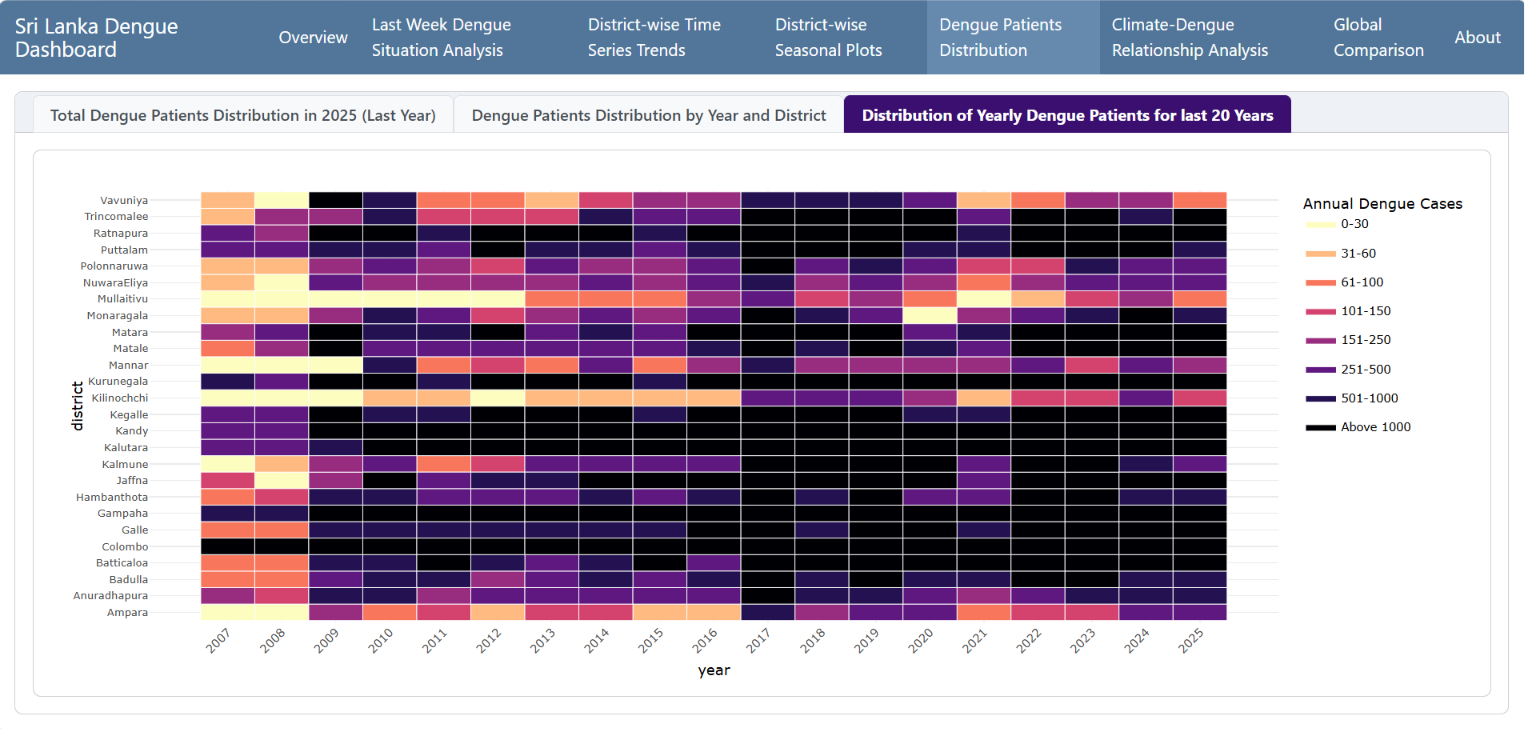}

}

\subcaption{\label{fig-tab5-3}}

\end{minipage}%

\caption{\label{fig-tab5}Screenshot of Tab 5 (Dengue Patients
Distribution)}

\end{figure}%

Figure~\ref{fig-tab5}(c) shows that the Western Province, particularly
the districts of Colombo, Gampaha, and Kalutara, has consistently
reported the highest dengue burden in the country. As shown in
Figure~\ref{fig-tab5}(b), these districts are geographically very small.
However, the Western Province serves as Sri Lanka's administrative,
commercial, and economic hub, attracting a large number of people to
live and work in these areas.

Consequently, the population density in these districts is exceptionally
high. In addition, rapid urbanization, extensive construction
activities, environmental conditions favourable for mosquito breeding,
and increased human mobility have further contributed to the high
transmission of dengue in the region. Figure~\ref{fig-tab5}(a) clearly
illustrates the hierarchical distribution of dengue cases across all
districts in 2025 using a treemap.

\begin{figure}[H]

\begin{minipage}[t]{0.50\linewidth}

\centering{

\includegraphics[width=0.98\linewidth,height=\textheight,keepaspectratio]{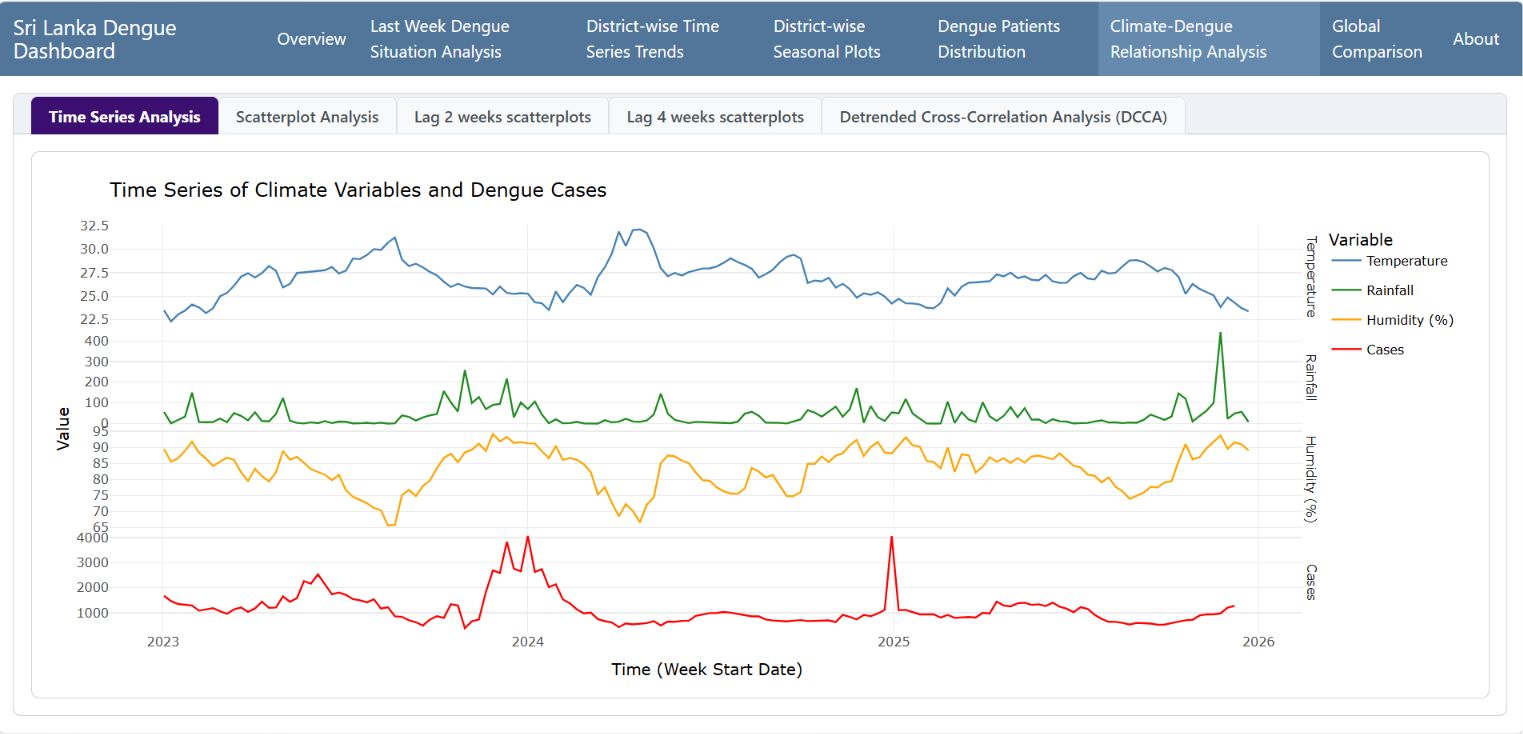}

}

\subcaption{\label{fig-tab6-1}}

\end{minipage}%
\begin{minipage}[t]{0.50\linewidth}

\centering{

\includegraphics[width=0.98\linewidth,height=\textheight,keepaspectratio]{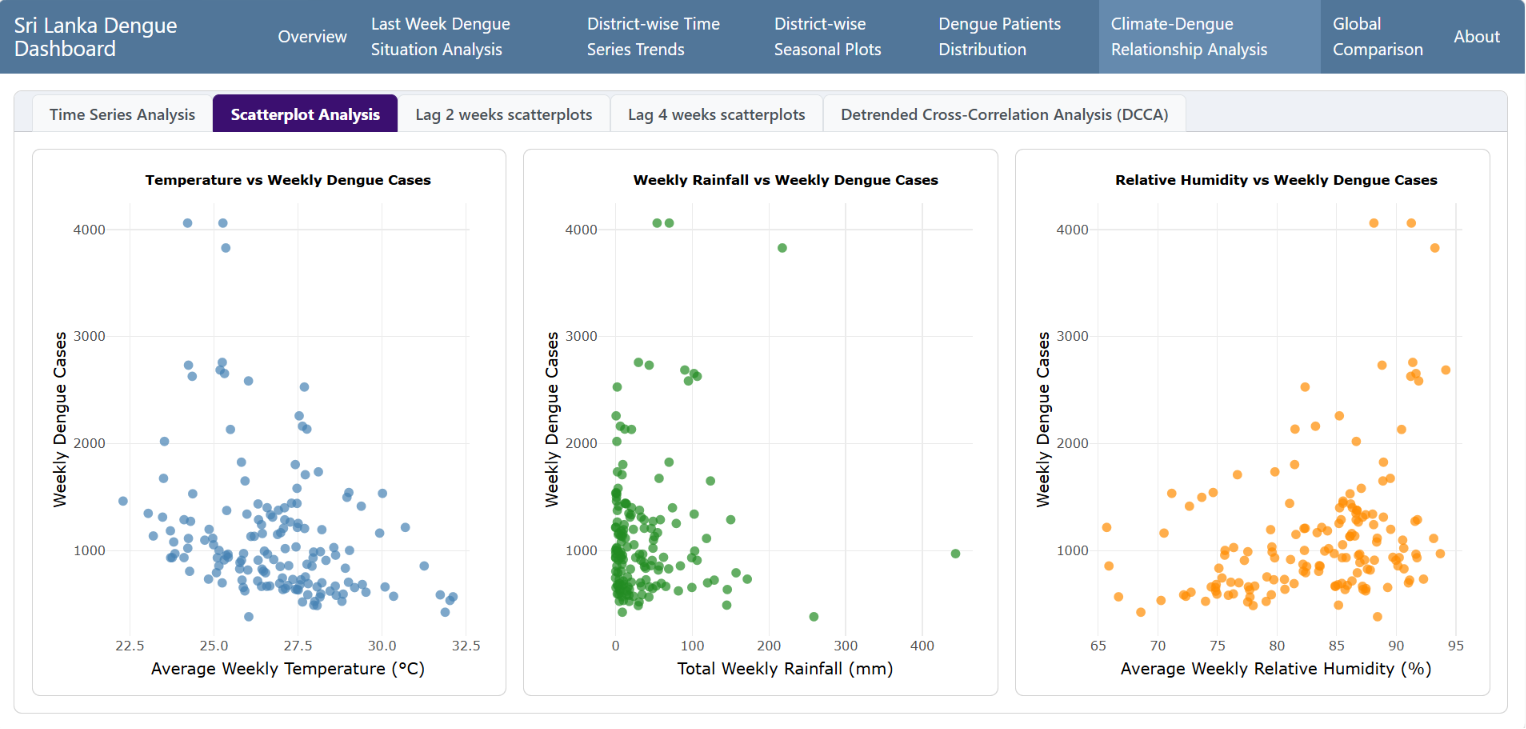}

}

\subcaption{\label{fig-tab6-2}}

\end{minipage}%
\newline
\begin{minipage}[t]{0.50\linewidth}

\centering{

\includegraphics[width=0.98\linewidth,height=\textheight,keepaspectratio]{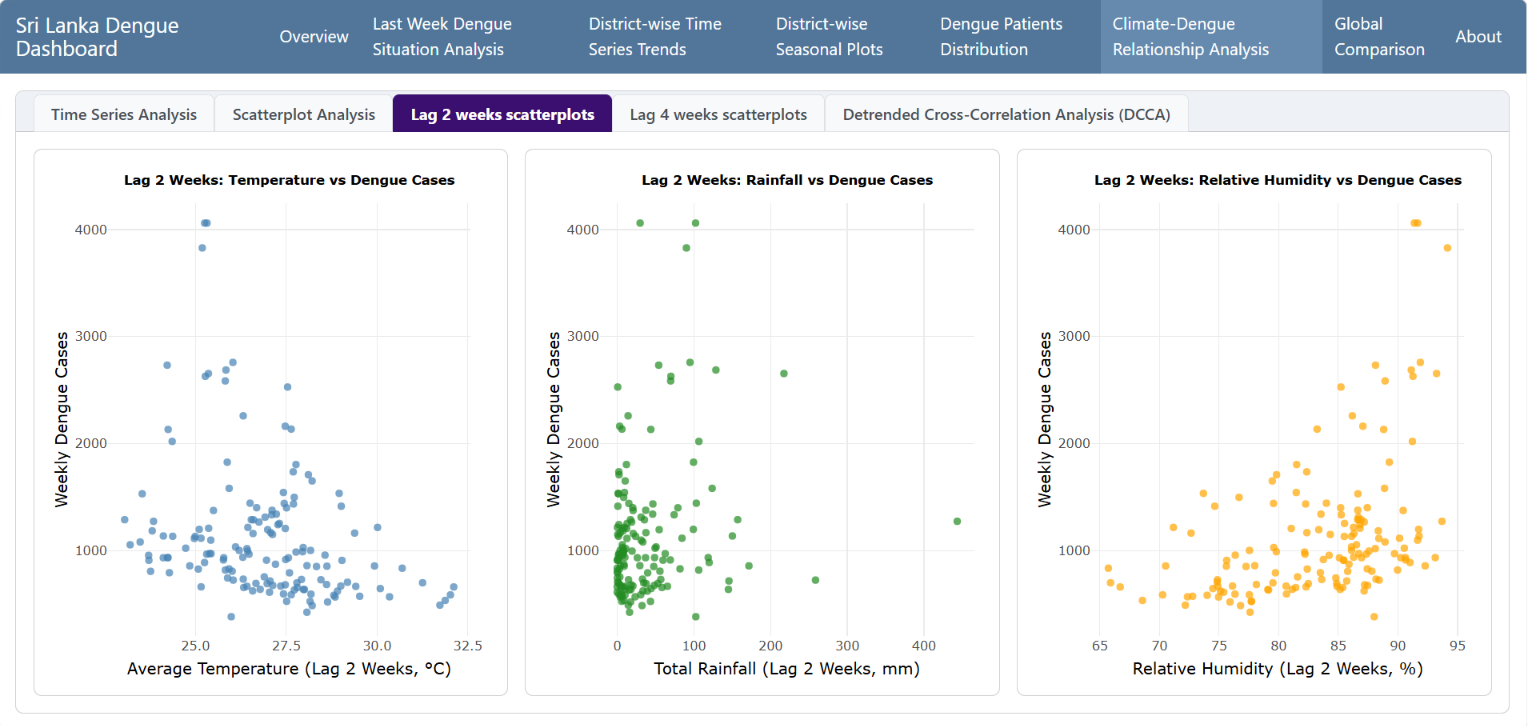}

}

\subcaption{\label{fig-tab6-3}}

\end{minipage}%
\begin{minipage}[t]{0.50\linewidth}

\centering{

\includegraphics[width=0.98\linewidth,height=\textheight,keepaspectratio]{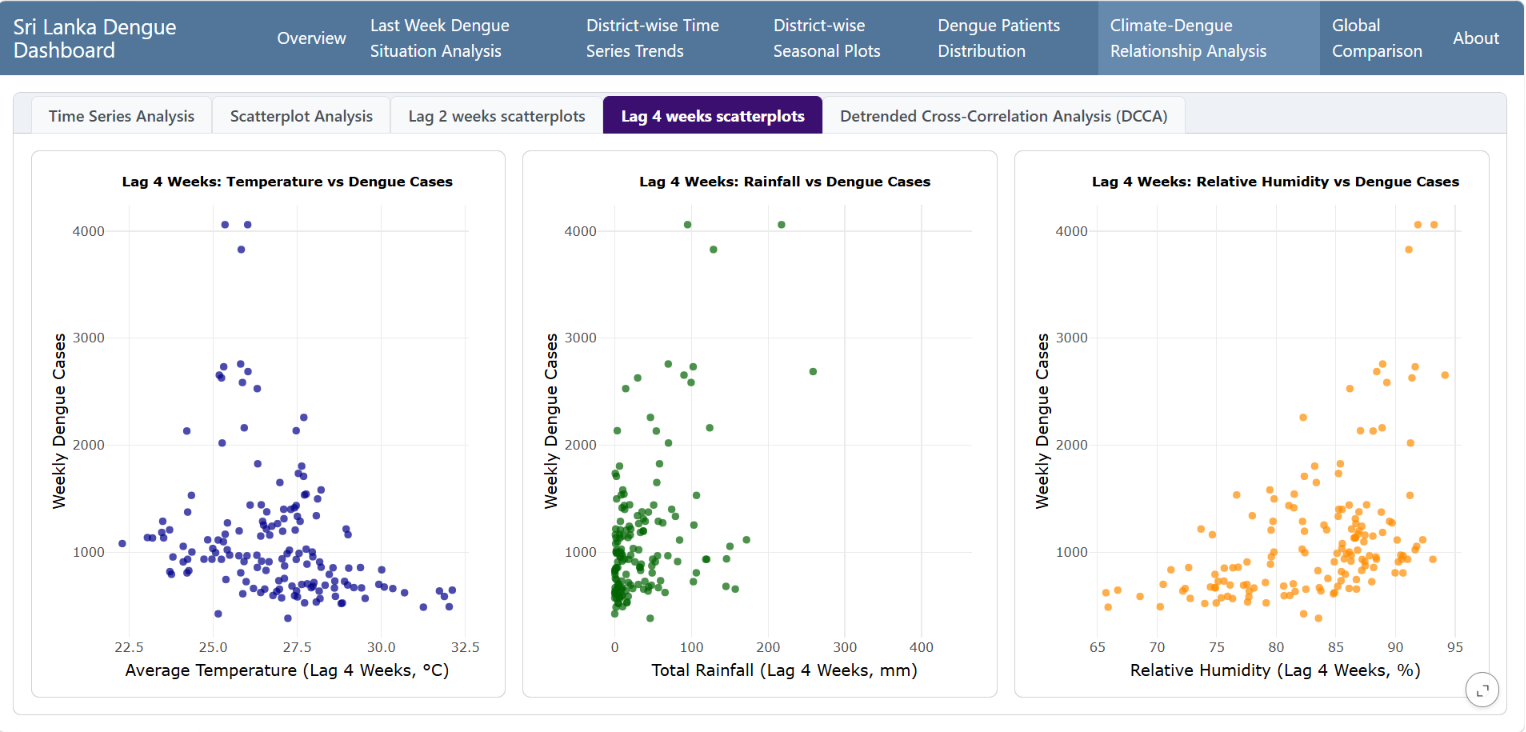}

}

\subcaption{\label{fig-tab6-4}}

\end{minipage}%
\newline
\begin{minipage}[t]{0.50\linewidth}

\centering{

\includegraphics[width=0.98\linewidth,height=\textheight,keepaspectratio]{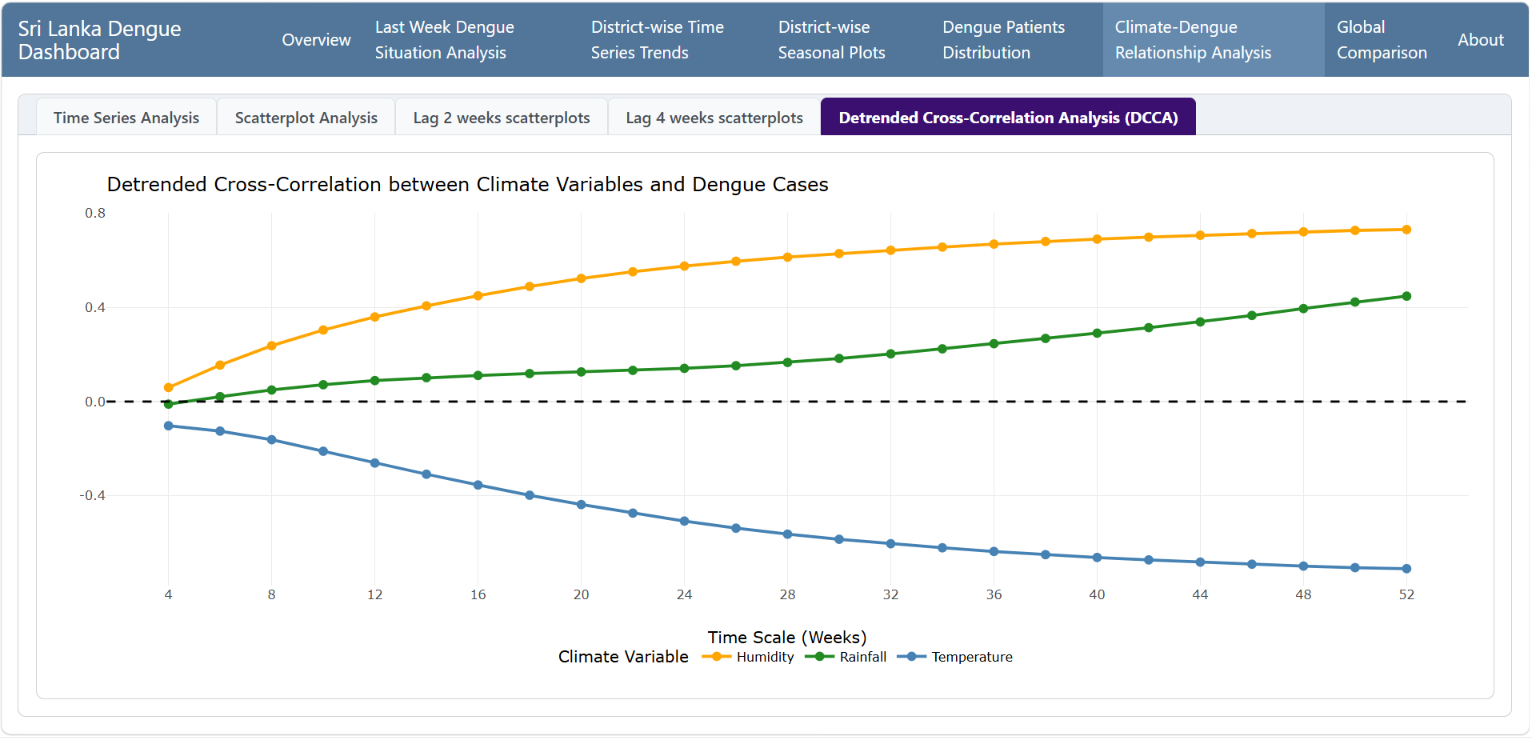}

}

\subcaption{\label{fig-tab6-5}}

\end{minipage}%

\caption{\label{fig-tab6}Screenshot of Tab 6 (Climate-Dengue
Relationship Analysis)}

\end{figure}%

Figure~\ref{fig-tab6}(a) shows a clearly visible relationship between
dengue cases and climatic variables, particularly relative humidity,
temperature, and rainfall. According to the figure, dengue cases
increase when relative humidity is high and temperature is low. During
periods with high dengue cases, rainfall is also higher than during
other periods.

When Figure~\ref{fig-tab6}(b), Figure~\ref{fig-tab6}(c), and
Figure~\ref{fig-tab6}(d) are examined, the above-mentioned lag effect
can be clearly seen. Among them, the scatter plots in
Figure~\ref{fig-tab6}(b) at a lag of 4 weeks show that the points are
more closely clustered and move in the same direction. This indicates a
positive relationship between relative humidity and dengue cases and a
negative relationship between temperature and dengue cases. However, the
relationship between rainfall and dengue cases is not as strong.

However, scatter plots and Pearson correlation coefficients are not very
suitable methods for quantifying the strength and direction of
relationships in time series data. Therefore, DCCA was used for this
purpose. According to the DCCA results shown in
Figure~\ref{fig-tab6}(e), there is no strong relationship in the short
run. However, in the long run, there is a strong positive relationship
between relative humidity and dengue cases, while temperature has a
strong negative relationship with dengue cases. The relationship between
rainfall and dengue cases is moderate in the long run.

\begin{figure}[H]

\begin{minipage}[t]{0.50\linewidth}

\centering{

\includegraphics[width=0.98\linewidth,height=\textheight,keepaspectratio]{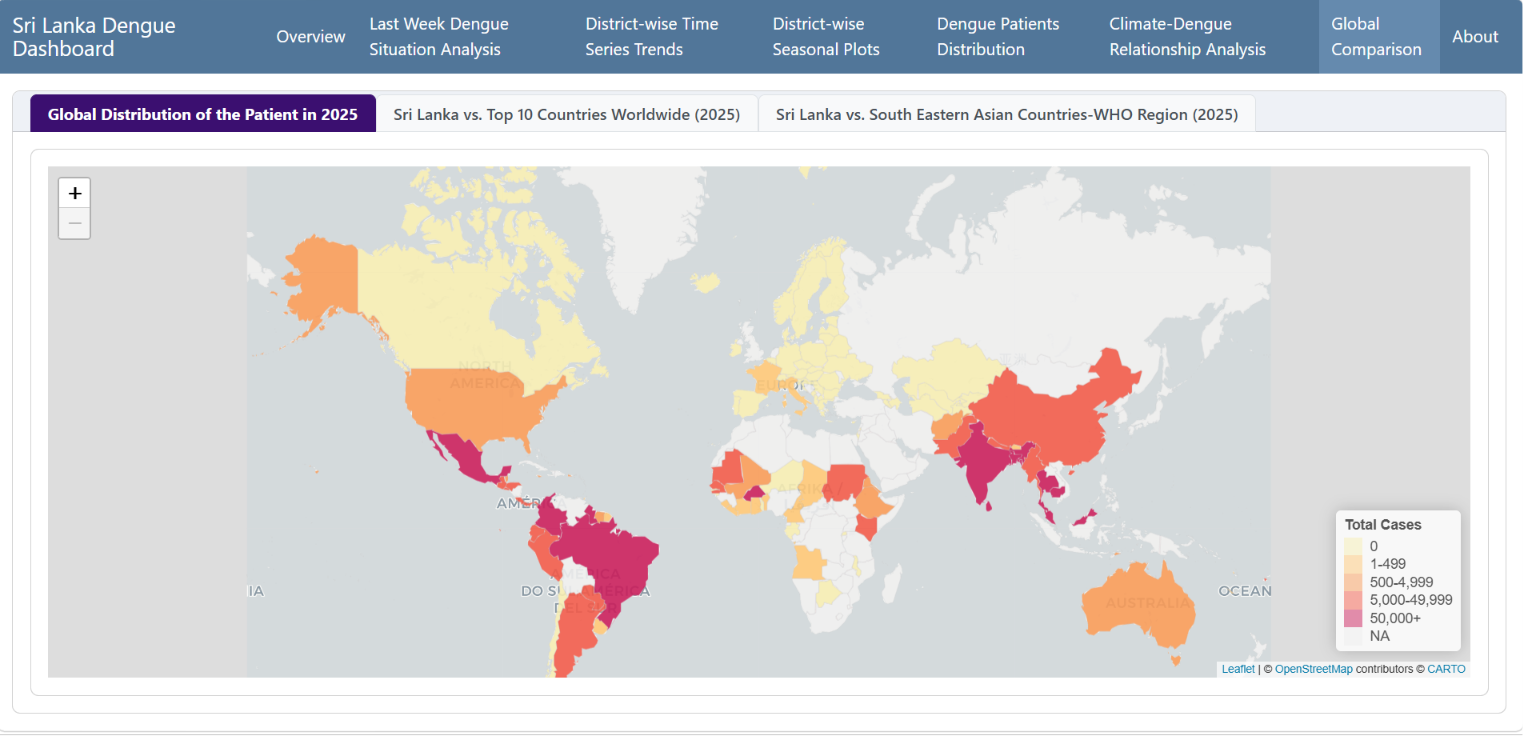}

}

\subcaption{\label{fig-tab7-1}}

\end{minipage}%
\begin{minipage}[t]{0.50\linewidth}

\centering{

\includegraphics[width=0.98\linewidth,height=\textheight,keepaspectratio]{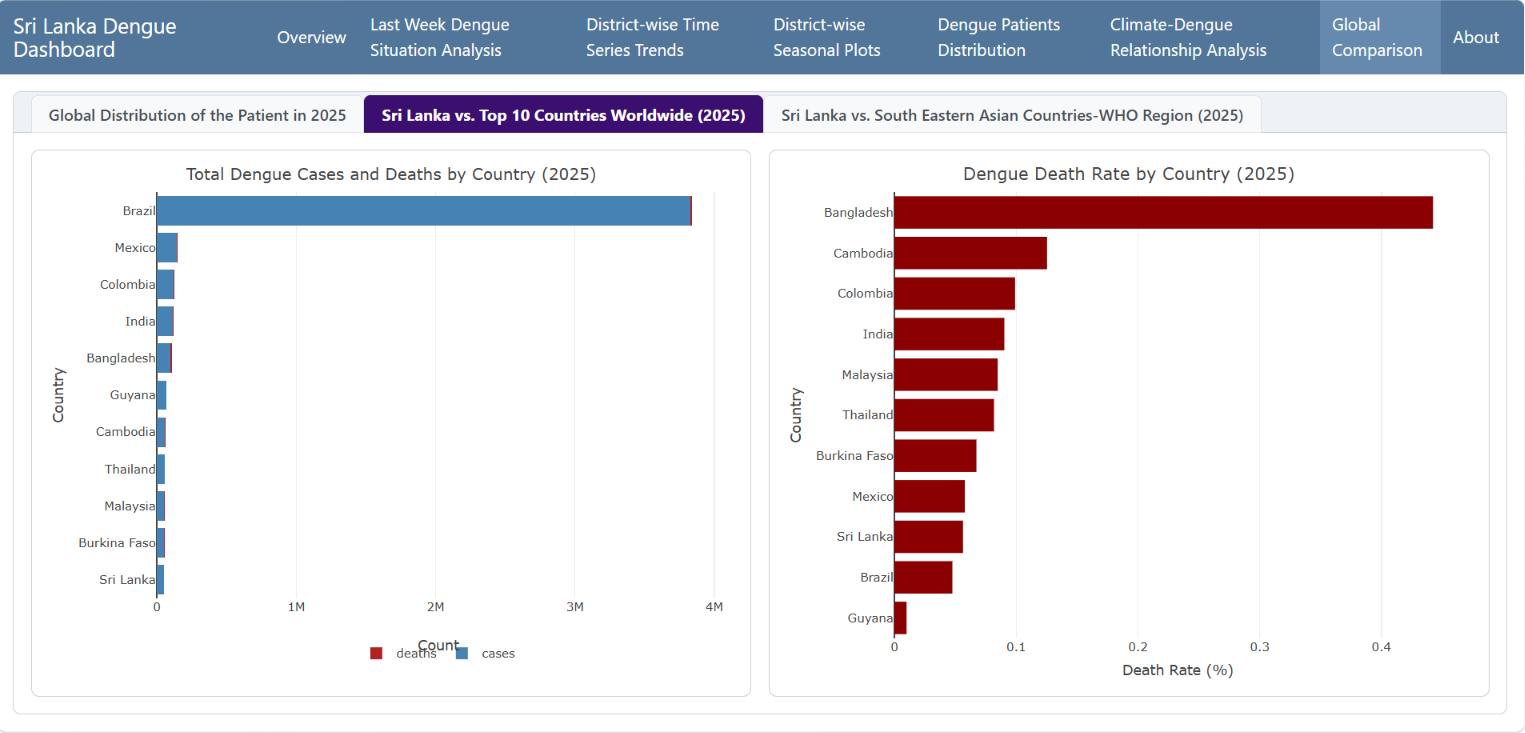}

}

\subcaption{\label{fig-tab7-2}}

\end{minipage}%
\newline
\begin{minipage}[t]{0.50\linewidth}

\centering{

\includegraphics[width=0.98\linewidth,height=\textheight,keepaspectratio]{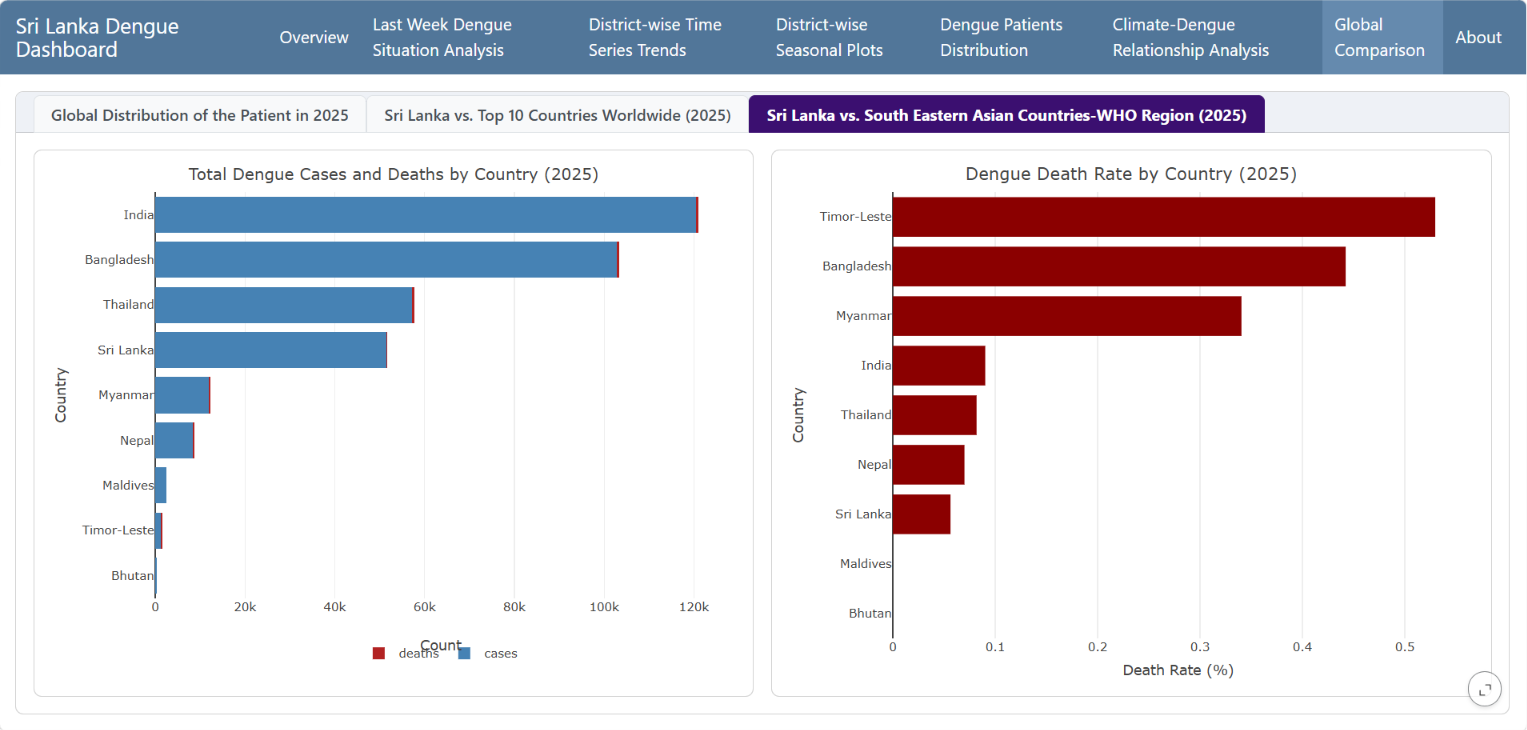}

}

\subcaption{\label{fig-tab7-3}}

\end{minipage}%

\caption{\label{fig-tab7}Screenshot of Tab 7 (Global Comparison)}

\end{figure}%

These findings may be highly useful for policy-making purposes and for
promoting personal precautionary activities. For example, when a period
of high relative humidity is observed for a certain duration, people can
be made aware that there may be an increased risk of dengue
transmission. Based on such early indications, health authorities can
organize public awareness programs, strengthen healthcare facilities,
and ensure the availability of necessary medical resources. At the
individual level, people can take preventive measures such as cleaning
their surroundings to eliminate mosquito breeding sites, using mosquito
repellents, and adopting other protective practices to reduce the risk
of dengue infection.

\begin{figure}[H]

\centering{

\includegraphics[width=5.11in,height=\textheight,keepaspectratio]{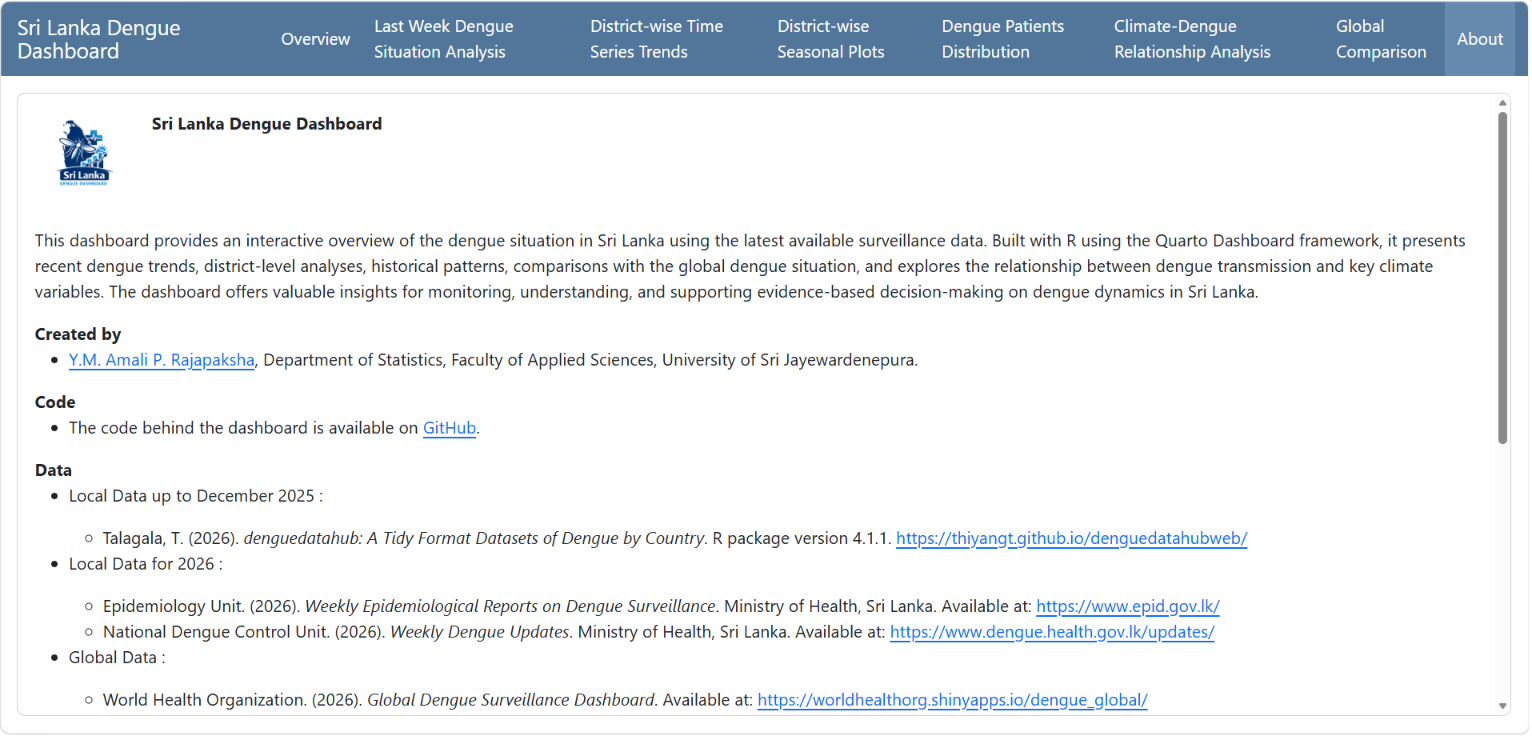}

}

\caption{\label{fig-tab8}Screenshot of Tab 8 (About)}

\end{figure}%

Figure~\ref{fig-tab7}(a) highlights the geographical variation of dengue
cases worldwide in 2025, while Figure~\ref{fig-tab7}(b) compares the
global situation with that of Sri Lanka. According to the figure,
although Sri Lanka has a lower number of dengue cases, its mortality is
comparatively higher, even exceeding that of some of the top 10
countries. However, when considering the South-East Asian countries, Sri
Lanka performs better than several regional peers, suggesting
comparatively lower mortality. This is shown in
Figure~\ref{fig-tab7}(c).

\section{6 Discussion}\label{discussion}

Most of the dashboards reviewed in this study have used bar charts, pie
charts, and line charts to present dengue cases and temporal trends.
However, modern visualization techniques such as tree maps and heatmaps
have not been widely used in these dashboards, whereas these techniques
have been incorporated into our dashboard to provide additional
perspectives for understanding dengue patterns. Among the reviewed
dashboards, only one dashboard provides daily updates. Similarly, our
dashboard does not currently support automatic updates and is developed
based on weekly dengue data rather than daily data. Therefore, a
potential direction for future research is to extend this work by
developing a daily, automatically updating dengue surveillance dashboard
for Sri Lanka.

This dashboard is fully reproducible, and all data sources and codes
used for its development are publicly available at
https://doi.org/10.5281/zenodo.21533747 (Rajapaksha, 2026).

\section*{References}\label{references}
\addcontentsline{toc}{section}{References}

\protect\phantomsection\label{refs}

\end{document}